\documentclass[a4paper,11pt]{article} 
\usepackage{jcappub}
\usepackage[latin1]{inputenc}
\usepackage{dsfont} 
\usepackage[amssymb]{SIunits} 
\usepackage[english]{babel}
\usepackage{bm}
\usepackage{dsfont} 

\newcommand{\eref}[1]{eq.~(\ref{#1})}

\newcommand{\sref}[1]{section~\ref{#1}}

\title{The Liouville equation for flavour evolution of  neutrinos and neutrino wave packets}

\author{Rasmus Sloth Lundkvist Hansen}
\author{and Alexei Yu. Smirnov}

\affiliation{Max-Planck-Institut f\"ur Kernphysik,\\ Saupfercheckweg 1, 69117 Heidelberg, Germany}
\emailAdd{rasmus@mpi-hd.mpg.de}
\emailAdd{smirnov@mpi-hd.mpg.de}

\abstract{
We consider several aspects related to the form, derivation and applications of the 
Liouville equation (LE) for flavour evolution of neutrinos. 
To take into account the quantum nature of neutrinos 
we derive the evolution equation for the 
matrix of densities using wave packets 
instead of Wigner functions. The obtained equation differs from the standard LE
by an additional term which is proportional to the difference of group 
velocities. We show that this term describes 
loss of the propagation coherence in the system.  
In absence of inelastic collisions, the LE can be  
reduced to a single derivative equation over a trajectory coordinate. 
Additional time and spacial dependence 
may steam from initial (production) conditions. The transition from single neutrino 
evolution to the evolution of a neutrino gas is considered.
}

\begin{document}

\maketitle
\flushbottom

\section{Introduction}

A field of propagating neutrinos inside a supernova can be described 
by the matrix of densities in the flavour space, $n (t,\mathbf{x},\mathbf{v},E)$, 
which depends on the space-time 
coordinates $(\mathbf{x},t)$, velocity $\mathbf{v}$ and energy $E$.  
The evolution of the matrix of densities in space and time is governed by the Liouville equation (see e.g. \cite{Strack:2005ux})
\begin{equation}
  \label{eq:liouville}
  i \mathcal{D} n (\mathbf{x},t,\mathbf{v},E) = 
[H(\mathbf{x},t,\mathbf{v},E), n(\mathbf{x},t,\mathbf{v},E)],
\end{equation}
where 
\begin{equation}
  \label{eq:LO}
  \mathcal{D} \equiv \partial_t + \mathbf{v} \cdot \partial_{\mathbf{x}}
\end{equation}
is the Liouville operator\footnote{We will not address the term involving $\partial_{\mathbf{x}} H \partial_{\mathbf{p}}$ in this work.} and 
$H(\mathbf{x},t,\mathbf{v},E)$ is the Hamiltonian 
that depends on properties of the background.  
The velocity matrix $\mathbf{v}$ is usually taken to be proportional 
to the unit matrix although its flavour dependence has previously been noticed \cite{Strack:2005ux,Sigl:1992fn}.

Eq.~(\ref{eq:liouville}) has been derived in a number of 
papers~\cite{Strack:2005ux, Sigl:1992fn, Rudzskii:1990, Sirera:1999, 
Yamada:2000za, Yamamoto:2003yy, Cardall:2007zw, Vlasenko:2013fja}
and most of the derivations are based on a presentation of  
the density matrix in terms of the Wigner function~\cite{Wigner:1932eb}.  
The latter  is a specific Fourier transform of the neutrino field bilinear. 
Furthermore, most of the derivations rely on formalism 
of statistical mechanics \cite{Akhiezer:1981}.

There is a number of questions related to the form, derivation and applications 
of \eref{eq:liouville} for supernova neutrinos.

\begin{enumerate}
\item [1)] Using the Wigner function for the derivation of the LE from first 
principles. This function allows 
to reconcile  the classical nature of the Liouville equation 
obtained from conservation 
of phase space with the quantum nature of propagating neutrinos. 
Indeed, in order to derive the Liouville equation, 
the simultaneous treatment of the neutrino momentum and the neutrino coordinate 
is needed~\cite{Akhiezer:1981}. 
Such a treatment of a quantum system can be obtained using the Wigner functions. 
However, the Wigner functions come with the price that in an inhomogeneous background,
they can attain negative values, and this would also propagate to 
quantities such as the matrix of densities.

\item[2)] Origin of two separate derivatives 
over time and spatial coordinates in the equation.
In some recent studies  the derivatives 
in space and time in the Liouville equation 
are treated independently~\cite{Dasgupta:2015iia,Capozzi:2016oyk}. 
One concern regarding these studies  is that the locality 
of the solution could be violated 
and that information could be transferred superluminally from one space point 
to another~\cite{Abbar:2015fwa}. 

\item[3)] The correspondence of \eref{eq:liouville} to
the evolution equation of the individual neutrino.
Indeed, in the absence of inelastic collisions
we can in principle consider evolution of the
individual neutrino moving along a certain trajectory
(and not an ensemble). 
For a single neutrino, two derivatives in \eref{eq:liouville} are reduced
to the one over the coordinate 
along the trajectory. Then a gas of neutrinos is just the
system of independently evolved neutrinos, and the density is the sum over
neutrinos in a given unit volume.
In general, for a gas, $\mathbf{x}$ and $t$ are independent since different
values of $(\mathbf{x},t)$ correspond to different neutrinos. For a single neutrino
$\mathbf{x}$ and $t$ are connected by $\mathbf{x} = \mathbf{v} t$ up to
uncertainty principle.

\item[4)] The form and nature of  the matrix ${\mathbf  v}$ in \eref{eq:liouville}.
Notice that in quantum theory the velocity $\mathbf{v}$ follows from the solution
of the equation rather than appears in the equation itself.
However if the inelastic collisions are neglected
and one can consider propagation of single neutrinos,
then ${\mathbf  v}$  should have an interpretation in terms of velocities
of the individual neutrinos.
In the case of two mixed neutrinos we can speak about
two group velocities. Using the averaged velocity in the LE, so that
$\mathbf{v} = \mathbf{v}_0 \mathds{1}$, should be justified. 
Indeed,  the difference of velocities in
${\mathbf  v}$ produces the term $\Delta {\mathbf  v} {\mathbf p} \approx \Delta m^2/2|\mathbf{p}|$
(in vacuum), which is comparable to the term with the Hamiltonian
(after subtraction of the part proportional to the unit matrix).
\end{enumerate}

In this paper we will address these four groups of questions. 
Instead of Wigner functions we propose to use wave packets 
which explicitly encode the uncertainty relation, and so, allow to reconcile 
the momentum and coordinate description.
Our method is the following: We  reconstruct the solutions 
of the evolution equation for the wave function using physics arguments and known asymptotic behaviour for certain known cases.
Then using this solution we find the equation which it satisfies, 
and finally we obtain the corresponding equation for the density matrix.
In a sense we solve the inverted problem here. 
We confront the obtained equation with the standard Liouville equation. 
Our analysis is applied to the cases of the uniform medium and a medium with 
adiabatically varying parameters where the eigenstates of the evolution equation are well defined.
The case of a non-adiabatic density change and neutrino self-interactions will be considered elsewhere~\cite{Hansen:2016}.
This allows us to clarify the origin of two (time and coordinate) 
derivatives in LE.  We study how and under which conditions 
the evolution equation is reduced from the full Liouville form with 
three spacial and one temporal derivative 
to a differential equation with only one derivative.
Our treatment allows also to clarify form and meaning of the matrix of velocities.
The transition 
from the single particle description to the description of a gas 
is important for this discussion. 
Using a modified evolution equation 
for a single neutrino we derive the equation for the matrix of densities. 

The paper is organised as follows. 
In \sref{sec:pointlike}  we will show that  the standard Liouville equation 
is satisfied for a system  of point-like particles. 
In \sref{sec:wavepackets} we will find the differential equation that corresponds 
to the  wave packet description of propagating neutrinos
and confront it with the usual Liouville equation. 
The obtained equation  contains an additional  term, as compared to the LE, proportional to 
the difference of the group velocities. 
In \sref{sec:decoherence} we study the physical meaning of this term and consider explicit examples of 
Gaussian and exponential forms of the wave packet. 
In \sref{sec:gas} we derive the evolution equations for a neutrino gas.   
Our conclusions are presented in \sref{sec:conclusions}.

\section{Liouville equation for point-like particles}
\label{sec:pointlike}

Let us consider a gas of freely propagating point-like neutrinos.  
A neutrino emitted at $(\mathbf{x}_0,t_0)$ is described by a wave function
\begin{equation}
  \label{eq:wavefunc}\nonumber
  \psi (\tau,\mathbf{x}_0,t_0),
\end{equation}
where $\tau$ is the time elapsed since emission, and $\psi$ is a vector in flavour space, $\psi^T = (\psi_1,\psi_2,\dots)$.
Treated as a point-like particle,
the neutrino has a well determined velocity $\mathbf{v}=(v_x,v_y,v_z)$. 
We assume here the same $\mathbf{v}$ for all eigenstates  of propagation
thus ignoring decoherence. 
For point-like particles with different velocities the coherence will be lost immediately.

The neutrino flavour evolution is described by  a Schr\"odinger-like equation
\begin{equation}
  i \frac{d}{d\tau} \psi(\tau,\mathbf{x}_0,t_0) = 
H(\mathbf{x},t) \psi(\tau,\mathbf{x}_0,t_0) ,
\label{eq:evoleq}
\end{equation}
where the Hamiltonian $H$ is in general a function of the \emph{local coordinates}
$(\mathbf{x},t)$.

For the individual neutrino, we define the density matrix in flavour space as 
\begin{equation}
\nonumber
\rho \equiv \psi \psi^\dagger.  
\end{equation}
Using  (\ref{eq:evoleq}) it is straightforward to check that 
$\rho$ satisfies the equation 
\begin{equation}
i \frac{d}{d\tau} \rho (\tau) = [H(\tau),\rho(\tau)]. 
\label{eq:evoleq1}
\end{equation}

Let us show that being expressed in terms of $(\mathbf{x},t)$, 
the density  matrix $\rho$ satisfies the Liouville equation. 
For an individual neutrino, the local coordinates are related to the coordinates of emission: \begin{equation}
\mathbf{x}= \mathbf{x}_0 + \mathbf{v}\tau, \quad t = t_0 + \tau.  
\label{eq:localvar}
\end{equation}
Therefore expressed in terms of the local coordinates, the density matrix 
can be presented as 
$$
\rho = \rho \left(\mathbf{x}(\tau), t(\tau)\right). 
$$
Consequently, 
\begin{equation}
\nonumber
\frac{d\rho}{d\tau} 
= \frac{\partial \rho}{\partial \mathbf{x}} \frac{\partial \mathbf{x}}{\partial \tau} 
+ \frac{\partial \rho}{\partial t} \frac{\partial t}{\partial \tau}
=  \mathbf{v} \frac{\partial \rho}{\partial \mathbf{x}} + 
\frac{\partial \rho}{\partial t}
= \mathcal{D} \rho. 
\end{equation}
Inserting this derivative into (\ref{eq:evoleq1}) we obtain 
\begin{equation}
\label{eq:L-rho2}\nonumber
  i \mathcal{D} \rho(\mathbf{x},t)  =  
\frac{\partial \rho (\mathbf{x},t)}{\partial \tau} = [H(t) , \rho(\mathbf{x},t)],
\end{equation}
which coincides with the standard form of the Liouville equation. 
The reason of this result is that due to the point-like nature, 
the local  variables $\mathbf{x}$ and $t$ are related 
according to (\ref{eq:localvar}), 
and essentially depend on a single  variable $\tau$, so that 
$$
\mathcal{D} \rho = \frac{d\rho}{ d\tau}. 
$$
The Liouville operator is nothing but the total derivative over time.

\section{Wave packets}
\label{sec:wavepackets}

Essentially, the consideration in \sref{sec:pointlike} was a classical one. 
In particular, because it uses exact velocities and spatial coordinates 
simultaneously for the description.
Here, instead of the Wigner function, we will use the wave packet 
description to account for the quantum nature of the neutrinos. 

In what follows we will reconstruct the solution of the evolution equation for the density 
matrix using wave packets.
As for the point-like neutrinos, the wave function in position space, 
$\psi^T = (\psi_1, \psi_2 ...)$ with $\psi_j$ being the components in flavour space
is given by the Fourier transform:
\begin{equation}
\label{eq:spacewavepsi}\nonumber
\psi(\mathbf{x},t) = \int \frac{d \mathbf{p}}{(2\pi)^3}  \; 
e^{i \mathbf{p} \cdot \mathbf{x}} \phi(\mathbf{p},t).
\end{equation}
Here the evolution of $\phi^T = (\phi_1,\phi_2,\dots)$ is  determined by the Schr\"odinger equation
\begin{equation}
\label{eq:Sphi}\nonumber
  i \frac{d\phi(\mathbf{p},t)}{dt} = H(\mathbf{p}, t) \phi(\mathbf{p},t).
\end{equation}
In a non-uniform medium, the Hamiltonian depends on time $t$ along the trajectory.

It is convenient to solve the evolution equation in the basis of 
eigenstates of the instantaneous Hamiltonian. 
In this basis 
\begin{equation}
\label{eq:ham}\nonumber
H(\mathbf{p},t)  = {\rm diag}[H_1(\mathbf{p},t),~ H_2(\mathbf{p},t), ...], 
\end{equation}
where $H_j$ are the eigenvalues of the Hamiltonian. 
In a uniform medium or a medium with adiabatically changing density the 
evolution equation splits into separate equations for the eigenstates $\phi_j(\mathbf{p},t)$:  
\begin{equation}
\label{eq:Sphij}
  i \frac{d\phi_j(\mathbf{p},t)}{dt} = H_j(\mathbf{p}, t) \phi_j(\mathbf{p},t).
\end{equation}
The integration of (\ref{eq:Sphij}) gives for the $j$th  component:
\begin{equation}
\label{eq:phii} 
\phi_j(\mathbf{p},t) =  f_j(\mathbf{p}) e^{-i \int^t_0 dt' H_j(\mathbf{p},t')},
\end{equation}
where $f_j(\mathbf{p})$ is the spectral function in momentum space.

In what follows we will neglect the difference in shape of $f_j(\mathbf{p})$ for 
different $j$: $f_j(\mathbf{p}) = A_j f(\mathbf{p})$, where $A_j$ are the numerical
coefficients given by mixing parameters at the production. (The shape difference is of the 
order $\Delta m^2/E^2$ and it does not increase with time, in contrast to the phases.) 
Additionally, we assume that in momentum space the spectral function peaks around the 
characteristic momentum in the packet, $\mathbf{p}_w$: $f(\mathbf{p}) = f(\mathbf{p} - \mathbf{p}_w)$ with width $\sigma_p \ll |\mathbf{p}_w|$. 

Thus, for a uniform medium, the $j$th component equals
\begin{equation}
\label{eq:spacewavef}
\psi_j(\mathbf{x},t)  =  A_j \int \frac{d \mathbf{p}}{(2\pi)^3}  \;
e^{i \mathbf{p} \cdot \mathbf{x} - i \int^t_0 dt' H_j(\mathbf{p},t')} f(\mathbf{p} - \mathbf{p}_w). 
\end{equation}
We can further specify $\psi_j$ by performing an expansion of the eigenvalues $H_j(\mathbf{p},t)$ around
the characteristic momentum $\mathbf{p}_w$: 
\begin{equation}
  \label{eq:Happrox}
  \begin{aligned}
  H_j(\mathbf{p},t) & = H_j(\mathbf{p}_w,t) + \left.\frac{\partial H_j(\mathbf{p},t)}{\partial \mathbf{p}}\right|_{\mathbf{p} =
\mathbf{p}_w} (\mathbf{p}-\mathbf{p}_w) + \mathcal{O}((\mathbf{p}-\mathbf{p}_w)^2)\\
  &\approx H_j(\mathbf{p}_w,t) + \mathbf{v}_{wj}(t) (\mathbf{p}-\mathbf{p}_w),
  \end{aligned}
\end{equation}
where 
\begin{equation}
  \label{eq:Vj}
  \mathbf{v}_{wj} = \left.\frac{\partial H_j(\mathbf{p},t)}{\partial \mathbf{p}}\right|_{\mathbf{p}
= \mathbf{p}_w}.
\end{equation}
Inserting the expression (\ref{eq:Happrox}) into (\ref{eq:spacewavef}), we obtain 
(see also \cite{Akhmedov:2009rb})
\begin{equation}
\label{eq:spacewave1}
\psi_j(\mathbf{x},t)  =    
A_j e^{i \mathbf{p}_w \cdot \mathbf{x} - i \int^t_0 dt' H_j(\mathbf{p}_w,t')}
 g(\mathbf{x} - {\textstyle \int^t_0} dt' \mathbf{v}_{wj}(t')), 
\end{equation}
where 
\begin{equation}
\label{eq:shapef}\nonumber
g[\mathbf{x} - {\textstyle \int^t_0} dt' \mathbf{v}_{wj}(t') ] = 
\int \frac{d \mathbf{p}}{(2\pi)^3}  \;
e^{i (\mathbf{p} - \mathbf{p}_w) \cdot (\mathbf{x} - \int^t_0 dt'\mathbf{v}_{wj}(t'))} 
f(\mathbf{p} - \mathbf{p}_w) 
\end{equation}
is the shape factor which  propagates 
with velocity $\mathbf{v}_{wj}$. 
This shows that  $\mathbf{v}_{wj}$ defined in (\ref{eq:Vj})  
can be identified with the group velocity of the $j$'th eigenstate. 
In the following the dependence of $H$ on $\mathbf{p}_w$ will be implicit.

To obtain a form of the evolution equation as close as possible to the standard Liouville equation, we will use the average velocity 
\begin{equation}
  \label{eq:v0}\nonumber
  \mathbf{v}_0(t) \equiv \frac{1}{n}(\mathbf{v}_{w1}(t) + \mathbf{v}_{w2}(t) + \dots + \mathbf{v}_{wn}(t))  
\end{equation}
in the Liouville operator.
With this, the Liouville operator acting on (\ref{eq:spacewave1}) gives
\begin{equation}
  \label{eq:Lpsin}
  i (\partial_t + \mathbf{v}_0(t) \cdot \partial_{\mathbf{x}} ) \psi(\mathbf{x},t) = (H(t) - \mathbf{v}_0(t) \cdot \mathbf{p}_w) \psi(\mathbf{x},t) - i \mathbf{Y}(t) \mathbf{G}(\mathbf{x},t) \psi(\mathbf{x},t).
\end{equation}
Here 
\begin{equation}
  \label{eq:Vexp}
  \mathbf{Y}(t) = \mathbf{V}_w(t) - \mathbf{v}_0(t)\mathds{1}, \quad \mathbf{V}_w(t)  = {\rm diag}(\mathbf{v}_{w1}(t), \mathbf{v}_{w2}(t), ... ),
\end{equation}
so that $\mathbf{Y}$ is the traceless matrix proportional to the differences of group velocities and
\begin{equation}
\label{eq:gfact}
\mathbf{G}(\mathbf{x},t) \equiv {\rm diag}(\mathbf{G}_1(\mathbf{x},t),  \mathbf{G}_2(\mathbf{x},t), ...  ), 
\quad \mathbf{G}_j(\mathbf{x},t) \equiv \frac{ \partial_{\mathbf{x}} g(\mathbf{x}-\int_0^t dt' \mathbf{v}_{wj}(t')) }{ 
g(\mathbf{x}-\int_0^t dt' \mathbf{v}_{wj}(t'))}. 
\end{equation}
It is important to notice that $H$, $\mathbf{Y}$ and $\mathbf{G}$ are diagonal matrices, and the matrix $\mathbf{Y}$ can be written as
\begin{equation}
\label{eq:tracelessY}
\mathbf{Y}(t) = 
\left.\frac{\partial H_{\mathrm{traceless}}(\mathbf{p},t)}{\partial \mathbf{p}} 
\right|_{\mathbf{p} =\mathbf{p}_w} .
\end{equation}
Furthermore, $\mathbf{Y}$ is much smaller than $|\mathbf{v}_0|$.
Explicitly, in the two neutrino case and with propagation in vacuum, we find
\begin{equation}
\mathbf{V}_w = 
\hat{\mathbf{p}}_w
\left[\left(1 - \frac{\bar{m}^2_{21}}{2 | \mathbf{p}_w|^2} \right) \mathds{1}
+
\frac{\Delta m^2_{21}}{4 |\mathbf{p}_w|^2} \sigma_3
    \right],  
\label{eq:velo}
\end{equation}
where $\hat{\mathbf{p}}_w = \mathbf{p}_w/|\mathbf{p}_w|$, $\bar{m}^2_{21} \equiv (m^2_{2} +  m^2_{1})/2$, and $\sigma_3 \equiv \mathrm{diag}(1, -1)$ is the third Pauli matrix. 
$\mathbf{Y}$ should be identified with the second flavour asymmetric and traceless part in  \eref{eq:velo}. 

Computing the time derivative of  $\rho \equiv\psi \psi^\dagger$ and using  
\eref{eq:Lpsin}, we obtain the evolution equation for the density matrix:  
\begin{equation}
  \label{eq:Lrhon}
  i(\partial_t + \mathbf{v}_0(t)\cdot\partial_{\mathbf{x}}) \rho(\mathbf{x},t) = [H(t),\rho(\mathbf{x},t)] + \mathcal{C}(\mathbf{x},t),
\end{equation}
where
\begin{equation}
  \label{eq:C}
   \mathcal{C}(\mathbf{x},t) \equiv -i \left[\mathbf{Y}(t) \mathbf{G}(\mathbf{x},t) \rho(\mathbf{x},t) + \rho(\mathbf{x},t) \mathbf{G}^*(\mathbf{x},t) \mathbf{Y}(t)\right] 
\end{equation}
is the new term in comparison to the standard Liouville equation (\ref{eq:liouville}).
Being scalar in the position space, $\mathcal{C}$ is a matrix in the space of eigenstates. 

Since $\mathbf{Y}$ and $\mathbf{G}$ are diagonal and $\mathbf{G}$ is real in realistic examples (see \sref{sec:decoherence}), the correction term (\ref{eq:C}) can be written as 
the anticommutator
\begin{equation}
  \label{eq:C2}
\mathcal{C}(\mathbf{x},t) = - i \{ \mathbf{Y}(t) \mathbf{G}(\mathbf{x},t), \;\rho(\mathbf{x},t) \}.
\end{equation}
Consequently, the full evolution equation for the density matrix (\ref{eq:Lrhon}) becomes
\begin{equation}   
  \label{eq:Lrhon2}
  i(\partial_t + \mathbf{v}_0\cdot\partial_{\mathbf{x}}) \rho = [H,\rho]
- i \{\mathbf{Y}\mathbf{G}, \rho \},
\end{equation}
where the dependencies on $\mathbf{x}$ and $t$ have been left out for brevity.
Thus, in terms of $\mathbf{G}$ we obtain a closed equation for 
the density matrix. $\mathbf{G}$ is determined by the production 
process and therefore can be considered as an external factor (see below). 
Since both $\mathbf{G}$ and $\mathbf{Y}$ are real,  the correction 
$\mathcal{C}$ is imaginary for real values of $\rho$, which means that it 
describes effects of  absorption and/or loss of coherence. 
Eq.~(\ref{eq:Lrhon2}) is one of  the main results of our paper, and it 
takes into account the non-trivial matrix  character of $\mathbf{V}_w$.

In what follows, we will study properties of $\mathcal{C}$ and clarify its physical meaning.
Notice that $\mathcal{C}$ is proportional to the flavour 
asymmetric matrix of velocities 
$\mathbf{Y}$. If velocities are equal for all eigenstates, then $\mathbf{Y} = 0$, and 
consequently $\mathcal{C} = 0$. So, deviations from the 
standard Liouville equation is related to the presence of 
different group velocities in the system. 

If $\mathcal{C}$ is small compared to $[H,\rho]$, the standard Liouville equation 
provides a good description of the evolution. 
Notice that the dominant flavour invariant component of $H$ 
gives a vanishing commutator and does not contribute to the evolution. 
The flavour non-symmetric component of $H$ is due to the difference of 
eigenvalues which depends on masses and background potentials.
Therefore the smallness of $\mathbf{Y}$ in comparison to $H$ cannot be related to the
smallness of the neutrino mass.  

Let us estimate $\mathcal{C}$. 
The general expression in (\ref{eq:tracelessY}) leads to the simple estimate 
$|\mathbf{Y}| \sim H_{\mathrm{traceless}}/|\mathbf{p}_w|$. 
In turn, $\mathbf{G}$ can be estimated using the size of the shape factor $\sigma_x$: $\mathbf{G} \sim 1/\sigma_x \sim \sigma_p$.
Consequently, 
\begin{equation}
\label{eq:Cest}
\mathcal{C} \sim  |\mathbf{G}| |\mathbf{Y}| \rho
\sim \frac{\sigma_p}{|\mathbf{p}_w|} H_{\mathrm{traceless}} \rho. 
\end{equation}
Thus, the correction $\mathcal{C}$ is suppressed in comparison to the main term in (\ref{eq:Lrhon}) by the ratio $\sigma_p/|\mathbf{p}_w| \ll 1$,
and for a wave packet which is narrow in momentum space,   
the Liouville equation is a good approximation. 
However, even in this case the correction from (\ref{eq:C}) can become significant 
over time, as we will see in \sref{sec:decoherence}.\\

The matrix nature of $\mathbf{V}$ was considered 
in various  derivations of the Liouville equation.
In particular, in~\cite{Sigl:1992fn} the following  equation has been obtained:\footnote{
We still ignore the term $\frac{i}{2}\{\partial_{\mathbf{p}}\rho,\partial_{\mathbf{x}}H\}$ 
in \cite{Sigl:1992fn}.}
\begin{equation}
  \label{eq:Lanticommutator}\nonumber
  i\partial_t\rho + \tfrac{i}{2} \{ \mathbf{V},\partial_{\mathbf{x}}\rho \} = [H,\rho].
\end{equation}
Using the expression for the velocity matrix (\ref{eq:Vexp}),
it can be rewritten as 
\begin{equation}
\label{eq:Lanticommutator1}
  i(\partial_t + \mathbf{v}_0\cdot\partial_{\mathbf{x}})\rho  = 
[H,\rho]  - \tfrac{i}{2} \{ \mathbf{Y}, \partial_{\mathbf{x}}\rho \}.  
\end{equation}
Let us compare (\ref{eq:Lanticommutator1}) with our result in (\ref{eq:Lrhon2}). 
For the wave function in \eref{eq:spacewave1} we find 
$$
\partial_{\mathbf{x}}\rho = \mathbf{G}\rho +  \rho \mathbf{G}^*. 
$$
With this, the last term in (\ref{eq:Lanticommutator1}) (analogy of our $\mathcal{C}$) 
becomes
\begin{equation}
 \nonumber
\mathcal{C'} \equiv 
- \tfrac{i}{2} \left[\{\mathbf{Y},  \mathbf{G}\rho \} + 
\{\mathbf{Y}, \rho \mathbf{G}^* \} \right] ,
\end{equation}
which is still different from  our expression for $\mathcal{C}$. 
The two expressions coincide  provided that
\begin{enumerate}
\item[1)] $\mathbf{G}$ is real,   $\mathbf{G} = \mathbf{G}^*$, 
which, indeed, is satisfied for the examples we will discuss.
\item[2)] $\mathbf{G}$ is flavour-symmetric,  $\mathbf{G} \propto  \mathds{1}$, that  is,   
$\mathbf{G}$  commutes with $\rho$. 
\end{enumerate} 
Under these conditions 
\begin{equation}
\label{eq:C3}
\mathcal{C} = \mathcal{C'} = - i \mathbf{G} \{\mathbf{Y}, \rho \}.  
\end{equation}

According to our consideration, condition 2), and consequently \eref{eq:C3}, are not satisfied due to different group velocities even with the assumption of universal spectral functions $f$. As we will show in \sref{sec:decoherence}, the difference of group velocities leads to separation 
of the wave packets and loss of propagation coherence. 

Finally, we remark that there is also an alternative route to 
the evolution \eref{eq:Lrhon}. 
Using \eref{eq:spacewave1}
we can reconstruct the density matrix in the
position space immediately: The diagonal elements are
\begin{equation}
\rho_{jj} =  A_j^2 g^2(\mathbf{x} -  {\textstyle \int^t_0} dt' \mathbf{v}_{wj}(t')) ,
\label{eq:diag-x}
\end{equation}
and the off-diagonal elements equal
\begin{equation}
\rho_{jk} =  A_j A_k \; g[\mathbf{x} - {\textstyle \int^t_0} dt' \mathbf{v}_{wj}(t')] \;
g[\mathbf{x} - {\textstyle \int^t_0} dt' \mathbf{v}_{wk}(t')] \;
e^{-i \int^t_0 dt' (H_j(t') -  H_k(t'))} .
\label{eq:offdiag-x}
\end{equation}
Applying the Liouville operator on the matrix formed by (\ref{eq:diag-x}) and (\ref{eq:offdiag-x}), we obtain (\ref{eq:Lrhon}).

\section{Decoherence}
\label{sec:decoherence}

Here we  study the physical effects of the correction $\mathcal{C}$.
To simplify calculations we consider first propagation in a uniform medium or in vacuum.
In this case $H$ does not depend on $t$ and consequently $\int^t_0 dt' H_j(t') = H_j t$ and 
$\int^t_0 dt' \mathbf{v}_{wj}(t') = \mathbf{v}_{wj} t$.
Considering two neutrinos allows $\mathbf{Y}$ to be written
\begin{equation}
\label{eq:tracelessY2}
\mathbf{Y} = 
- \frac{1}{2} \frac{\partial (H_2 - H_1)}{\partial \mathbf{p}} \sigma_3 = 
\frac{1}{2}\Delta \mathbf{v} \sigma_3 ,
\end{equation}
where $\Delta \mathbf{v} \equiv \mathbf{v}_{w1} - \mathbf{v}_{w2}$. In vacuum, $\Delta \mathbf{v} = \hat{\mathbf{p}}_w \Delta m_{21}^2 / (2 |\mathbf{p}_w|^2)$, 
which can also be seen from (\ref{eq:velo}).
Using the definition (\ref{eq:C2}) and $\mathbf{Y}$ (\ref{eq:tracelessY2}), we obtain 
\begin{equation}
  \label{eq:Cexpl}\nonumber
\mathcal{C} 
=  - i \frac{ \Delta \mathbf{v}}{2}  \{ \sigma_3 \mathbf{G},\; \rho \}.
\end{equation}
In terms of components of the density matrix, $\mathcal{C}$ can be rewritten as
\begin{equation}
  \label{eq:Ctwo-expl}
  \mathcal{C} = - i \Delta \mathbf{v}
\begin{bmatrix}
     \mathbf{G}_1 \rho_{11} & \frac{1}{2}(\mathbf{G}_1 - \mathbf{G}_2) \rho_{12} \\ 
     \frac{1}{2} (\mathbf{G}_1 - \mathbf{G}_2) \rho_{21}  & - \mathbf{G}_2  \rho_{22}
    \end{bmatrix}. 
\end{equation}
In what follows we find $\mathcal{C}$ for specific shape factors. 

For the Gaussian shape factor 
\begin{equation}
  \label{eq:gGauss}
  g(\mathbf{x}-\mathbf{v}_{wj}t) = \left(\tfrac{\sigma_p}{\sqrt{\pi}}\right)^{3/2} 
e^{-\frac{1}{2}(\mathbf{x}-\mathbf{v}_{wj}t)^2\sigma_p^2} ,
\end{equation}
(which also corresponds to a Gaussian spectral function $f(\mathbf{p})$)
using the definition (\ref{eq:gfact}), we obtain
\begin{equation}
  \label{eq:dgdxGauss}
\mathbf{G}_j = -\sigma_p^2 (\mathbf{x} - \mathbf{v}_{wj}\ t).
\end{equation}
As we assumed in \sref{sec:wavepackets} $\mathbf{G}$ is, indeed, real. 
Since $|\mathbf{x}-\mathbf{v}_{wj}t| \sim \sigma_x \sim 1/\sigma_p$, we have 
$|\mathbf{G}_j| \sim \sigma_p$.
This agrees well with the simple estimate used to obtain (\ref{eq:Cest}).

Inserting  expression (\ref{eq:dgdxGauss}) for $\mathbf{G}_j$ into (\ref{eq:Ctwo-expl})
we find
\begin{equation}
  \label{eq:Ctwostates}
  \mathcal{C} = i \Delta \mathbf{v}\sigma_p^2
\left( 
   \begin{bmatrix}
      (\mathbf{x}-\mathbf{v}_{w1} t) \rho_{11} & 0 \\ 0 & -(\mathbf{x}-\mathbf{v}_{w2} t) \rho_{22} 
    \end{bmatrix}
    - \frac{\Delta \mathbf{v}t}{2} 
    \begin{bmatrix}
      0 & \rho_{12} \\ \rho_{21} & 0
    \end{bmatrix}
    \right) .
\end{equation}
According to (\ref{eq:Ctwostates}),
\begin{enumerate}
\item[1)] $\mathcal{C} \propto \sigma_p^2 \sim 1/\sigma_x^2$, 
that is, the shorter the wave packet in the position space, the stronger the effect. 
\item[2)] $\mathcal{C} \propto \Delta \mathbf{v}$ --- the correction is proportional to the difference of group velocities. 
\end{enumerate}
From this, one can conclude that $\mathcal{C}$  describes the effect 
of loss of coherence due to separation of the wave packets. 
This can also be inferred from the matrix structure of $\mathcal{C}$,
where in (\ref{eq:Ctwostates}) 
the diagonal part separates the densities (wave packets) 
of the two eigenstates in space. Indeed,  $\mathbf{x} = \mathbf{v}_{wj} t$ 
determines the centres of the wave packet, so that 
$\mathbf{x} > \mathbf{v}_{wj} t$ is the front and 
$\mathbf{x} < \mathbf{v}_{wj} t$ is the back parts  
of the wave packet. For $\rho_{11}$ and the front part, we have $-i \mathcal{C} > 0$.
Therefore the $11$-term of the matrix increases the density distribution $\rho_{11}$ 
in front and decreases it in the back for $\mathbf{v}_{w1} > \mathbf{v}_{w2}$, while the $22$-term does the opposite. These changes are equivalent to a relative shift of the packets.

The second matrix in (\ref{eq:Ctwostates}) affects the off-diagonal part of the density matrix. 
Due to the sign and the factor of $(\Delta \mathbf{v})^2$, 
it leads to a decrease of the off-diagonal parts of $\rho$ in the uniform medium,
thus producing  decoherence. 

The physical meaning  of $\mathcal{C}$ can also be seen 
performing an integration of  \eref{eq:Lrhon} over $\mathbf{x}$:
\begin{equation}
  \label{eq:Ldx}
  i \int d^3\mathbf{x} (\partial_t + \mathbf{v}_0 \cdot \partial_{\mathbf{x}}) 
\rho(\mathbf{x},t) = 
\int d^3 \mathbf{x} [H,\rho(\mathbf{x},t)] + \int d^3\mathbf{x} \;\mathcal{C}(\mathbf{x},t).
\end{equation}
Let us define  $\rho(t) \equiv \int d^3\mathbf{x} \rho(\mathbf{x},t)$, 
and chose the integration volume large enough, so that $\rho(\mathbf{x},t)=0$ 
at the boundary of the volume.   Then the integral of the second term on the LHS 
(which can be reduced by the Stock theorem to the surface integral) vanishes.  
In a uniform medium  $H$ does not depend on $\mathbf{x}$, 
and consequently  \eref{eq:Ldx} gives
\begin{equation}
  \label{eq:Ldx1}
  i \partial_t \rho(t) = [H,\rho(t)] + \int d^3\mathbf{x} \;\mathcal{C}(\mathbf{x},t).
\end{equation}
Inserting expression (\ref{eq:Ctwostates}) into (\ref{eq:Ldx1}), 
we find that the diagonal matrix in (\ref{eq:Ctwostates}) integrates to zero because 
the density distribution of the $j$'th eigenstate 
is symmetric around $\mathbf{x} = \mathbf{v}_{wj} t$.
Consequently (\ref{eq:Ldx1}) becomes
\begin{equation}
  \label{eq:liouvilledecoh}
  i \partial_t \rho(t) = [H,\rho(t)] - i\frac{t}{2L_{\mathrm{coh}}^2}
  \begin{bmatrix}
      0 & \rho_{12}(t) \\ \rho_{21}(t) & 0    
  \end{bmatrix}.
\end{equation}
Here
\begin{equation}
  \nonumber
  L_{\mathrm{coh}} = \frac{1}{|\Delta \mathbf{v}| \sigma_p}
\end{equation}
is the coherence length and 
$\rho_{21}(t)$ and $\rho_{12}(t)$ are again the $\mathbf{x}$-integrated elements. 
This result can also be obtained by taking the explicit form of the wave packets 
from (\ref{eq:spacewave1}) and (\ref{eq:gGauss}) 
and calculating $\partial_t \int d\mathbf{x} \psi \psi^\dagger$.\\

As a second example, let us consider the case of exponential wave packets in one dimension 
which correspond to neutrinos produced in decays~\cite{Minakata:2012kg}.  
To make the packets continuous in whole space, we parametrise them as 
the limit $\sigma_f \rightarrow \infty$ of 
\begin{equation}
  \label{eq:exponential}
  g(x-v_{wj} t) =
  \begin{cases}
    \sqrt{2 \sigma_p} e^{\frac{1}{4}\sigma_p (x-v_{wj}t)}, \quad &\mathrm{for} \quad x < v_{wj}t , \\
    \sqrt{2 \sigma_p} e^{-\frac{1}{4} \sigma_f (x-v_{wj}t)}, \quad &\mathrm{for} \quad v_{wj}t < x ,
  \end{cases}
\end{equation}
$j = 1,~2$.
From this expression we obtain 
\begin{equation}
  \label{eq:Gexp}\nonumber
  G_j = G(x-v_{wj}t) =
  \begin{cases}
    \frac{1}{4}\sigma_p , \quad &\mathrm{for} \quad x < v_{wj} t ,\\
    -\frac{1}{4}\sigma_f , \quad &\mathrm{for} \quad v_{wj} t < x .
  \end{cases} 
\end{equation}
Then according to (\ref{eq:Ctwo-expl}) the correction 
$\mathcal{C}$ equals in assumption of $v_{w1} > v_{w2}$
\begin{equation}
  \label{eq:Cexp}\nonumber
  \mathcal{C} = -i \frac{\Delta v}{4}
  \begin{cases}
    \begin{bmatrix}
      \sigma_p \rho_{11} &&  0\\
      0 && - \sigma_p \rho_{22}
    \end{bmatrix}, &\mathrm{for}\; x < v_{w2} t, \vspace{5pt}\\
    \begin{bmatrix}
      \sigma_p \rho_{11} &&  \frac{1}{2} (\sigma_p + \sigma_f) \rho_{12}\\
      \frac{1}{2} (\sigma_p + \sigma_f) \rho_{21} && \sigma_f \rho_{22}
    \end{bmatrix}, &\mathrm{for}\; v_{w2} t < x < v_{w1} t, \vspace{5pt}\\
    \begin{bmatrix}
      - \sigma_f \rho_{11} &&  0\\
      0 && \sigma_f \rho_{22}
    \end{bmatrix}, &\mathrm{for}\; v_{w1} t < x .
  \end{cases}
\end{equation}
Notice that the off-diagonal terms of $\mathcal{C}$ are non-zero only in the second region between the two maxima of the wave packets. 
The coefficient in these terms, $\frac{1}{2}(\sigma_p + \sigma_f)$,  grows unrestrictedly  
when $\sigma_f \rightarrow \infty$. This is not a problem since according to 
\eref{eq:exponential} $\rho_{12}$ and $\rho_{21}$ decrease exponentially 
in the same limit.
However, it highlights that the evolution equations only make sense for a continuous $g$.

The evolution equation for the $x$-integrated density matrix is given in 
(\ref{eq:Ldx1}). The correction integral can be computed 
using the expression for  $\rho_{12}$ in (\ref{eq:offdiag-x}) 
with the explicit form of the shape factors in 
(\ref{eq:exponential}). In the second region, which gives the correction,   
we have 
\begin{equation}
\nonumber
\rho_{12}(x, t) = A_1 A_2 2 \sigma_p e^{\frac{1}{4}(\sigma_p - \sigma_f) x - \frac{1}{4}(\sigma_p v_{w1} - \sigma_f v_{w2})t} 
e^{i\Delta H}. 
\end{equation}
Its integration leads to 
\begin{equation}
  \label{eq:Cexpint1}\nonumber
  \int_{v_{w2} t}^{v_{w1} t } dx \mathcal{C}_{12} = 
- i A_1 A_2 \frac{\Delta v}{4} \sigma_p\frac{\sigma_p + \sigma_f}{\sigma_p-\sigma_f} 
(e^{-\frac{1}{4}\sigma_f \Delta vt} 
- e^{-\frac{1}{4}\sigma_p \Delta v t}) e^{- i(H_1 - H_2)t},
\end{equation}
and in the limit $\sigma_f \rightarrow \infty$, we obtain 
\begin{equation}
  \label{eq:Cexpint2}
  \int_{v_{w2} t}^{v_{w1} t } dx \mathcal{C}_{12} \rightarrow  
- i A_1 A_2 \frac{\Delta v \sigma_p}{4} e^{-\frac{1}{4}\sigma_p \Delta v t - i(H_1 - H_2)t}  .
\end{equation} 
Integration of  $\rho_{12}$ over the whole space (over all three regions) 
gives 
\begin{align}
\rho_{12}(t) &\equiv  \int_{-\infty}^{+\infty} dx \rho_{12}(x,t)\nonumber\\
&= A_1 A_2 \left[e^{-\frac{1}{4}\sigma_p \Delta v t} + 
\frac{2\sigma_p}{\sigma_p-\sigma_f} (e^{-\frac{1}{4}\sigma_f \Delta vt} - e^{-\frac{1}{4}\sigma_p \Delta vt} ) 
+ e^{-\frac{1}{4}\sigma_f \Delta vt}\right]e^{- i(H_1 - H_2)t}.  
  \label{eq:rhoexp}\nonumber
\end{align}
And in the limit $\sigma_f \rightarrow \infty$: 
\begin{equation}
\label{eq:rhoexpl}
\rho_{12}(t) \rightarrow  A_1 A_2 e^{-\frac{1}{4}\sigma_p \Delta v t - i(H_1 - H_2)t}.
\end{equation}
Using (\ref{eq:rhoexpl}) we can express the correction (\ref{eq:Cexpint2}) in terms of $\rho_{12}(t)$ in the limit $\sigma_f \rightarrow \infty$:
\begin{equation}
  \label{eq:Cexpint}
  \int_{-\infty}^\infty dx \mathcal{C} = - i \frac{1}{4 L_{\mathrm{coh}}}
  \begin{bmatrix}
    0 && \rho_{12}(t)\\
    \rho_{21}(t) && 0
  \end{bmatrix},
\end{equation}
which is also valid for $v_{w1}<v_{w2}$.
As in the Gaussian case, we see that the diagonal terms disappear 
as they must preserve probability.

The result in \eref{eq:Cexpint} differs from the result for the Gaussian
wave packet case in \eref{eq:liouvilledecoh} by a factor $2t/|L_{\mathrm{coh}}|$.
Since the effect of wave packet separation is significant at $t\sim L_{\mathrm{coh}}$, 
the two expressions give similar results, 
although the decoherence is less pronounced at early times for the Gaussian wave packet.
The transition region is sharper in time in the Gaussian case.
The correction term found in (\ref{eq:Cexpint}) agrees with the one used 
in \cite{Akhmedov:2014ssa}, the only difference is a factor due to the 
definitions of $\sigma_p$.

The decoherence can also be seen in the explicit form of the density matrix. The off-diagonal elements from (\ref{eq:offdiag-x}) integrated over $\mathbf{x}$:
$$
e^{-i (H_j -  H_k)t} 
A_j A_k \int_{-\infty}^{+\infty} d \mathbf{x}g(\mathbf{x} -\mathbf{v}_{wj} t) g(\mathbf{x} - \mathbf{v}_{wk} t),
$$ 
disappears with time due to decrease of the overlap 
of $g(\mathbf{x} -\mathbf{v}_{wj} t)$ and $g(\mathbf{x} - \mathbf{v}_{wk} t)$, 
as the consequence of different group velocities .  

In a non-uniform medium, $\Delta \mathbf{v}$ varies and furthermore changes sign  
close to the MSW resonance \cite{Mikheyev:1989dy}. 
Consequently, the eigenstates might decohere at high densities and later restore 
partially or completely the coherence at lower density, where the difference 
in group velocity $\Delta \mathbf{v} = \partial (H_1 - H_2)/\partial \mathbf{p}$
has changed sign. 
To take such effects into account, we need to take into account the dependence of 
$H$ on time, and therefore restore the integral over time of $H$ and 
$\Delta \mathbf{v}$. Carrying this change through the calculations
for Gaussian wave packets gives the space integrated evolution equation corresponding to \eref{eq:liouvilledecoh}:
\begin{equation}
  \label{eq:Lgausst}\nonumber
  i \partial_t \rho(t) = [ H(t), \rho(t) ] - i \frac{\sigma_p^2\Delta \mathbf{v}(t)}{2}
  \begin{bmatrix}
      0 & \rho_{12}(t) \\ \rho_{21}(t) & 0    
  \end{bmatrix}
\int_0^t dt' \Delta \mathbf{v}(t')   
.
\end{equation}
The presence of both $\Delta \mathbf{v}(t)$ and $\int_0^t dt' \Delta \mathbf{v}(t')$ allows the equation to describe both decoherence and the restoration of coherence when $\Delta \mathbf{v}$ changes sign, but $\int_0^t dt' \Delta \mathbf{v}(t')$ does not.

\section{The neutrino gas. From a single neutrino to an ensemble of neutrinos}
\label{sec:gas}

The consideration in the previous sections has been done essentially
for a single neutrino and the density matrix $\rho$
is simply the matrix representation of the neutrino flavour polarisation vector 
(see e.g.~\cite{Raffelt:1996wa}).
Here we present a generalisation of the analysis to the case of
a field of propagating neutrinos characterised by the matrix of densities
in flavour space $n ({\bf x}, t, {\bf v}, E)$.
(We can use ${\bf p}$ instead of ${\bf v},\, E$). 
In the absence of inelastic collisions, the matrix of densities
$n$ can be reconstructed from the density matrices of individual neutrinos
propagating along  certain trajectories determined by  ${\bf v}$. 
After obtaining $n$, we will find the equation it satisfies.

Let us start with the simplest case of a
flux of point-like neutrinos $F(t)$ emitted from a
source at ${\bf x} = 0$. We consider first one spacial dimension
and relatively small distances, $x \ll L_{coh}$, so that
the difference of group velocities can be neglected. 
For a point-like neutrino $\rho (x, t) = \rho (x)$
depends on the distance from the production point,
and satisfies the single derivative 
equation\footnote{Formally we could introduce the time dependence in $\rho(x)$ as
\begin{equation}
\nonumber
\frac{d\rho(x, t)}{dt} = \rho (x) \delta(t - x/v). 
\label{eq:xt}
\end{equation}
This density, indeed, satisfies the LE and leads to the same
equation (\ref{eq:md-1df}) for the matrix of densities $n$.}
\begin{equation}
i v \partial_x \rho (x) = [H, \rho (x)]. 
\label{eq:lesingle}
\end{equation}
It also satisfies the Liouville  equation, so that  $v \partial_x$ can be substituted
by $\mathcal{D}$ since  $\partial_t \rho (x) = 0$.
The density of neutrinos in the point $(x, t)$ is given by
\begin{equation}
n(x, t) = \frac{1}{v} F(t - x/v) \rho(x). 
\label{eq:md-1d}
\end{equation} 
It is straightforward to show that $n(x, t)$ satisfies the
standard Liouville equation. Acting on $n(x, t)$
by the Liouville operator with single velocity we obtain
\begin{equation}
i \mathcal{D} n(x, t) = \frac{i}{v} F(t - x/v) \partial_x\rho (x)
= \frac{1}{v} F(t - x/v) [H, \rho (x)] = [H, n (x, t)].
\label{eq:md-1df}
\end{equation}
Here in the second equality we used \eref{eq:lesingle}
for the derivative $\partial_x \rho(x)$ and
in the last equality --- the definition (\ref{eq:md-1d}). 
Notice that in (\ref{eq:md-1df}) the derivatives of $F$ cancel: 
\begin{equation}
\mathcal{D} F(t - x/v) = 0
\label{eq:vanish}
\end{equation}
due to the dependence on coordinates in the combination $(t - x/v)$.

For an ensemble of point-like neutrinos, the time dependence follows from
the production condition and it is independent of the
$x$-derivative. As a consequence of the point-like character of neutrinos, there is no restriction on the speed of time-variations. The flux factors out from the 
evolution.

Notice that in more than 1D the divergency of the flux in space
is taken into account by the LE  which implies
conservation of the flux in the complete phase space
which includes both coordinates and momenta.

In the three dimensional case neutrinos are
emitted from a surface fixed by the coordinates
${\bf x}_0$. In general, the flux can depend on the emission point
on this surface and on the direction determined by ${\bf v}$.
So, $F = F({\bf x}_0, t_0, \mathbf{v})$, and in terms of coordinates $({\bf x},t)$:
\begin{equation}
F = F\left(t - \frac{L(\mathbf{x})}{v}, {\bf x}_0({\bf x})\right),
\label{eq:fdependence }\nonumber
\end{equation}
where $v = |{\bf v}|$ and $L({\bf x}) = |\mathbf{x}-\mathbf{x}_0(\mathbf{x})|$. 

Acting by the 3D  Liouville operator (with universal
${\bf v}$) on the matrix of densities 
\begin{equation}
n({\bf x}, t) = \frac{1}{v} F\left({\bf x}_0({\bf x}), t - \frac{L(\mathbf{x})}{v}\right) \rho(L(\mathbf{x}))
\label{eq:densityin3d}\nonumber
\end{equation} 
we obtain
\begin{equation}
i \mathcal{D} n(\mathbf{x}, t) = i  \frac{1}{v} \left[ (\mathcal{D} F) \rho (L)
+ F \mathcal{D} \rho (L) \right], 
\label{eq:lo3d}
\end{equation}
where $\rho (L)=\rho(L(\mathbf{x}))$ is the density matrix of an individual neutrino
propagating along the trajectory with a baseline $L$ and satisfies the equation
\begin{equation}
i \mathcal{D} \rho (L) = i v \partial_L \rho (L) = [H, \rho (L)].
\label{eq:lesingle3}
\end{equation} 
Here we have taken into account that 
$$
\mathbf{v} \frac{\partial}{ \partial \mathbf{x}} = v \frac{\partial}{\partial L}. 
$$

The action of the Liouville operator  on $F$ gives 
\begin{equation}
\mathcal{D} F = {\bf v} \partial_{\bf x_0} F
\frac{\partial {\bf x_0} }{\partial {\bf x}}
= \sum_{j,k} v^{(j)} \frac{\partial x_0^{(k)}}{\partial x^{(j)}} 
\frac{\partial F}{\partial x_0^{(k)} }, 
\label{eq:dfdf}
\end{equation}
where differentiation of $t_0 = t - L/v$ vanishes, 
as in (\ref{eq:vanish}). The term in (\ref{eq:dfdf}) is due to 
dependence of the flux on coordinate of the emitting surface. 
Let us show that this term vanishes too. 
Given coordinates $(\mathbf{x}, \mathbf{v})$ determine the 
emission point $\mathbf{x}_0$ on the emission surface. 
If the surface is continuous and differentiable 
around $\mathbf{x}_0$, we can consider a plane tangent to the surface 
at $\mathbf{x}_0$, and take a small patch of the plane around $\mathbf{x}_0$.  
Then the surface is well approximated by the tangent plane. 
We can consider the system of coordinates $\mathbf{x} = (x^{(1)},~ x^{(2)},~ x^{(3)})$  
in which the tangent plane is at  $x_{0}^{(3)} = 0$ and $\mathbf{v} = (v^{(1)}, 0, v^{(3)})$
(that is, the trajectory of the neutrino is in the plane with $x^{(2)} = 0$). 
In this coordinate system 
$$
x_0^{(1)} = x^{(1)} - x^{(3)} \frac{v^{(1)}}{v^{(3)}}, ~~~ x_0^{(2)} = x^{(2)},~~~ x_0^{(3)} = 0,   
$$ 
and then immediately, 
$$
\sum_j v^{(j)}  \frac{\partial x_0^{(k)}}{\partial x^{(j)}} = 0~~~~~ k = 1,~2,~3. 
$$
Consequently, the RHS of (\ref{eq:dfdf}) is zero, $\mathcal{D} F = 0$,  and 
\eref{eq:lo3d} becomes 
\begin{equation}
i \mathcal{D} n(x, t) = i  \frac{1}{v} F \mathcal{D} \rho (L) .
\label{eq:lo3dd}\nonumber
\end{equation}
Finally using \eref{eq:lesingle3} we again arrive at the standard Liouville equation 
for $n$:
\begin{equation}
i \mathcal{D} n = [H, n].  
\label{eq:lo3df}\nonumber
\end{equation}

Let us consider the case of individual neutrinos
described by wave packets. 
Now three scales are relevant for the problem:
The width of the wave packet in position space, 
$\sigma_x$, the effective distance between the  centres of the wave packets of neighbouring neutrinos, $\lambda$, 
and the coarse graining scale, $\Delta x$, in which we will compute the number of neutrinos. 
Then the density is obtained dividing the number of neutrinos by $\Delta x$.
For simplicity we will consider one spatial dimension, 
so that the distance  between the centres of the wave packet equals $\lambda = v \Delta t$  
and $\Delta t$ is the averaged interval between emission of two sequential neutrinos.\footnote{
Since we are doing a coarse graining, we use the same $v$ for all the eigenstates.} 
Then the  matrix of densities is
\begin{equation}
  \label{eq:mod}
  n(x,t) =  \frac{1}{\Delta x}\sum_{j=-\infty}^{\infty} \int_{x}^{x+\Delta x} dx' \rho(x',t-t_j),
\end{equation}
where $t_j$ is the time  of emission of the $j$'th neutrino. 
We take $\Delta x \gg \lambda$, so that in the interval $\Delta x$ there are many neutrinos,
otherwise, the description of the gas is not different from the description 
of every particle independently.

Depending on $\sigma_x$, we have two different physics cases: 

1) If  $\Delta x \ll \sigma_x$, the density matrix is almost constant 
on the interval $[x,x+\Delta x]$,  it can be put out of the integral in (\ref{eq:mod}), 
and the integration is trivial leading to  
\begin{equation}
  \label{eq:mod_sxl1}
  n(x,t) =  \frac{1}{\Delta x}\sum_{j=-\infty}^{\infty} \Delta x \rho(x,t-t_j).
\end{equation}
The  number of wave packets per coarse graining scale 
equals $n_x = \Delta x/ \lambda$, so that $\Delta x = n_x \lambda = n_x v \Delta t$. 
Inserting this expression into (\ref{eq:mod_sxl1}) and substituting 
summation by integration 
$$
\sum_{j=-\infty}^{\infty} \Delta t  \rightarrow \int dt'
$$
we obtain
\begin{equation}
  \label{eq:mod_sxl2}\nonumber
  n(x,t) =  \frac{n_x}{\Delta x} v \int_{-\infty}^{\infty}  dt' \rho(x,t-t').
\end{equation}
Since the integral over $\rho$ is in an integral over shape factors 
with the form $g(x-v_{wj}t)$, it is very well approximated by
\begin{equation}
  \label{eq:mod_sxl3}
  n(x,t) =  \frac{n_x}{\Delta x} \int_{-\infty}^{\infty}  dx' \rho(x',x/v),
\end{equation}
where it was used that the shape factors give a significant 
contribution to the integral when $t' \approx t - x/v$.
The number of neutrinos per $\Delta x$ can also be expressed 
in terms of the flux as $F/v$. If the flux changes slowly with time
compared to $\Delta x/v$ this does not affect the integral, and
\begin{equation}
  \label{eq:mod_sxl4}
  n(x,t) =  \frac{F(t-x/v)}{v} \int_{-\infty}^{\infty}  dx' \rho(x',x/v).
\end{equation}
We notice that the expression in (\ref{eq:mod_sxl4}) has the same form as the $n$ for point-like particles in (\ref{eq:md-1d}). As a consequence, the derivation is very similar to the derivation for point-like particles leading to (\ref{eq:md-1df}) with the difference that (\ref{eq:lesingle}) should be replaced by
\begin{equation}
  \label{eq:lesingleWP}
  i v \partial_x \rho(x) = [H,\rho(x)] + \int dx' \mathcal{C}(x',x/v). 
\end{equation}
Eq.~(\ref{eq:lesingleWP}) is the 1D version of \eref{eq:Ldx1} with the substitution $t \rightarrow x/v$.
The evolution equation for  $n(x,t)$ in (\ref{eq:mod_sxl4}) is then
\begin{equation}
  \label{eq:WPgas}
  i \mathcal{D} n(x,t) = [H,n(x,t)] + \frac{F}{v}\int dx' \; \mathcal{C}(x',x/v) ,
\end{equation}
where $\mathcal{C}$ is given in (\ref{eq:C2}).
The last term describes the wave packet decoherence as it has already been 
discussed in \sref{sec:decoherence}.

2) In the case of a narrow  wave packet, $\lambda, \sigma_x \ll \Delta x$,   
the matrix of densities can be obtained from (\ref{eq:mod}):  
\begin{equation}
  \label{eq:mod_lsx}
  n(x,t) =  \frac{1}{\Delta x} n_x \int_{-\infty}^{\infty} dx' \rho(x',x/v).
\end{equation}
Here it was used that almost all of the wave packets are fully contained in the interval, 
so that the integration limits can be extended to infinity. 
The number of wave packets in the interval $\Delta x$ is $n_x$,
and the emission time for wave packet at $(x,t)$ is given by $t_j = t-x/v$ as we also used in (\ref{eq:mod_sxl3}). 
Since (\ref{eq:mod_lsx}) gives the same expression for $n(x,t)$ as in (\ref{eq:mod_sxl3}), the evolution equation for the matrix of densities also coincides with \eref{eq:WPgas}.

Thus, we have demonstrated that the basic form of 
the Liouville equation for the density matrix can be obtained without 
referring to the Wigner functions, and hence it is 
be independent of the associated peculiarities.

\section{Conclusions}
\label{sec:conclusions}

\hspace{\parindent}1) We study the validity and applications of the Liouville equation
for oscillating neutrinos in vacuum and in matter.
Inelastic interactions and neutrino self-interactions were neglected here.
In contrast to the previous studies we have accounted for the
quantum nature of neutrinos using explicitly the wave packet
description of the neutrino state avoiding the use of Wigner functions.

2) We have reconstructed
explicit solutions for the wave functions and then
the density matrix using known exact solutions.
Then we have found equations
which such a density matrix satisfies and confronted
them with the standard Liouville equation.\\
The obtained equation differs from the Liouville
equation by an additional term $\mathcal{C}$ which is related to the
presence of components with different group velocities
(essentially, with different eigenvalues of the Hamiltonian).
We show that $\mathcal{C}$ describes the effect of loss of 
propagation coherence due to separation of the wave packets.\\
The correction $\mathcal{C}$ depends on the specific shape
of the wave packet which, in turn, is determined by
production conditions. It is proportional to
$t/2L_{\mathrm{coh}}^2$ for Gaussian and to $1/4L_{\mathrm{coh}}$
for exponential shape factors.
With $\mathcal{C}$ we put certain information about the
initial conditions into the evolution equation.
This is consistent with the fact that
the density matrix itself is constructed
using the wave packets.

3) In the wave packet approach the matrix
${\mathbf v}$ equals the unit matrix
multiplied by the average group velocity
of the eigenstates in the system.
The difference of the group velocities
appears in the additional term $\mathcal{C}$
in the equation.

4) We show that the density matrix constructed
with wave packets, and thus explicitly
accounting for the quantum nature of neutrinos,
satisfies the modified Liouville equation
with only one additional term which describes
separation of the wave packets.
The latter is unavoidable in the wave packet picture.\\
This also implies that one can simply consider
the standard Liouville equation for the
plane waves with equal momenta
for all eigenstates and then perform integration over
the production spectrum and the detection energy
(momentum) resolution. In this case  the
density matrix depends on time only;
derivatives over coordinates and velocity are irrelevant.

5) Our approach allows to clarify the meaning of
derivatives over time and coordinates in the
Liouville operator.  For an individual 
point-like neutrino the density matrix depends
essentially on a single variable
(time or coordinate along the trajectory related by
the equality $L = v t$).
The Liouville equation is satisfied
trivially since it can be reduced to the
Schr\"odinger-like equation.

6) If inelastic collisions are absent, the field of
neutrinos can be described as the sum
over individual freely propagating neutrinos.
In this case additional dependence of the
matrix of densities on time and coordinate may follow from
dependence of the produced neutrino flux on time and
the point on the emission surface. We show that in this case
the  matrix of densities still satisfies the modified LE.
The additional dependence does not affect the evolution equation
and can be taken in to account as an external factor.

7) Additional dependence of time may follow from
dependence of the matter potential on time.
If this dependence is not very fast,
it simply corresponds to an adiabatically changing $H$.
The special case of very fast time dependence or non-adiabatic
variation of density  can be realised
in the presence of neutrino self-interactions.
This case will be considered elsewhere.

\acknowledgments

The authors would like to thank Evgeny Akhmedov for useful discussions. 
RSLH acknowledges support from the Alexander von Humboldt Foundation.

\end{document}